\documentclass[11pt]{article}
\usepackage{amssymb,latexsym,amsmath,amsbsy}
\usepackage[dvips]{graphicx}
\usepackage{young}  
\usepackage{cite}
\usepackage{pdfpages}
\usepackage{hyperref}
\DeclareMathOperator{\ch}{char}
\DeclareMathOperator{\sdim}{sdim}
\newcommand{\myatop}[2]{\genfrac{}{}{0pt}{}{#1}{#2}}

\begin{document}
\noindent
{\Large \bf
Extension of the $\mathfrak{osp}(m|n) \sim \mathfrak{so}(m-n)$ correspondence
to the \\[2mm]
infinite-dimensional chiral spinors and self dual tensors}\\[5mm] 
{\bf N.I.~Stoilova$^1$, J.~Thierry-Mieg$^2$ and J.\ Van der Jeugt$^3$}\\[2mm]
$^1$Institute for Nuclear Research and Nuclear Energy,
Boul.\ Tsarigradsko Chaussee 72, 1784 Sofia, Bulgaria  \\
$^2$NCBI, National Library of Medicine, National Institute of Health,
8600 Rockville Pike, Bethesda MD20894, USA  \\
$^3$Department of Applied Mathematics, Computer Science and Statistics, Ghent University,\\
Krijgslaan 281-S9, B-9000 Gent, Belgium \\[2mm]
E-mail: stoilova@inrne.bas.bg,  mieg@ncbi.nlm.nih.gov and Joris.VanderJeugt@UGent.be

\vskip 10mm

\noindent

\noindent
\addtolength{\parskip}{2mm}

\begin{abstract}
The spinor representations of the orthosymplectic Lie superalgebras $\mathfrak{osp}(m|n)$ are considered and constructed.
These are infinite-dimensional irreducible representations, of which the superdimension coincides with the dimension 
of the spinor representation of $\mathfrak{so}(m-n)$.
Next, we consider the self dual tensor representations of $\mathfrak{osp}(m|n)$ and their generalizations: 
these are also infinite-dimensional
and correspond to the highest irreducible component of the $p^{th}$ power of the spinor representation.
We determine the character of these representations, and deduce a superdimension formula.
From this, it follows that also for these representations the $\mathfrak{osp}(m|n) \sim \mathfrak{so}(m-n)$ correspondence holds.
\end{abstract}


\section{Introduction}
\label{sec1}

In space-time, space and time are not equivalent. The metric of $\mathfrak{so}(3,1)$,
$ds^2 = dx^2 + dy^2 + dz^2 - dt^2$, counts time as negative distance.
From the point of view of the photon travelling at speed of light between two distant 
galaxies, the proper time and distance are zero. This explains why the photon only
has two degrees of freedom. Although it is described classically by a 4-component vector
of $\mathfrak{so}(3,1)$, its two polarization states form an $\mathfrak{so}(2)$
vector, defined in the two-dimensional space transverse
to the direction of propagation.

In order to avoid the mathematical difficulties in the rigorous
formulation of quantum field theory in Minkowski space-time, one may
use the formulation in Euclidean space 
$ds^2 = dx^2 + dy^2 + dz^2 + d\tau^2$, obtained by performing
a so-called Wick rotation $\tau = i t$, which changes the signature of the
metric and effectively replaces $\mathfrak{so}(3,1)$ by $\mathfrak{so}(4)$.
But this raises a problem, since the photon would acquire 4 degrees of freedom. The solution, inspired by 
 Feynman~\cite{Feynman}, De Witt~\cite{DeWitt}, Faddeev and Popov~\cite{FadeevPopov}, and by so called
BRST symmetry~\cite{Becchi,BonoraTonin} (see also~\cite{Bonora} and the references therein),
 is to add two anticommuting
directions $({\overline \theta},\theta)$ and use a
Euclidean metric in $(x,y,z,\tau)$ space and anti-symmetric symplectic metric in $({\overline \theta},\theta)$ space:
$ ds^2 = dx^2 + dy^2 + dz^2 + d\tau^2 + d{\overline \theta} d\theta -  d\theta d{\overline \theta} $,
effectively replacing  $\mathfrak{so}(3,1)$ by $\mathfrak{osp}(4|2)$ thus, replacing Minkowski
space-time $\mathbb{R}^{3,1}$ by the superspace $\mathbb{R}^{4|2}$~\cite{Delbourgo}.  

Further, one uses the fact that the two
Grassmann variables $\theta, \bar\theta$ are compensating the extra
dimensions created by the Wick rotation, thus, recovering the
$\mathfrak{so}(2)$ invariant transverse physical space.  
In terms of representation theory, one could say that the 6-dimensional defining representation of $\mathfrak{osp}(4|2)$,
with 4 ordinary variables and 2 Grassmann variables, has superdimension $4-2$, coinciding with the
dimension of the defining representation of $\mathfrak{so}(2)$.
This is a manifestation of what is called the $\mathfrak{osp}(4|2) \sim \mathfrak{so}(2)$ correspondence.
More generally, the coincidence of superdimensions of $\mathfrak{osp}(m|n)$ representations with
dimensions of $\mathfrak{so}(m-n)$ representations is called the $\mathfrak{osp}(m|n) \sim \mathfrak{so}(m-n)$
correspondence.

To quote other examples of this phenomenon for mass zero gauge fields, the $g_{\mu \nu}$ graviton and the anti-symmetric 3-tensors of
$\mathfrak{so}(10,1)$ supergravity follow the same pattern~\cite{Cremmer}.
The number of degrees of freedom are respectively 44 and 84, 
equal to the respective dimensions of the metric and the 3-tensor of $\mathfrak{so}(9)$.
Their so-called ghost spectra are given by the corresponding tensors of $\mathfrak{osp}(11|2)$~\cite{Baulieu}.

In the early days of Lie superalgebra representations,
the $\mathfrak{osp}(m|n) \sim \mathfrak{so}(m-n) \sim \mathfrak{sp}(n-m)$
correspondence was established for various classes of representations~\cite{Green-Jarvis,Balantekin-Bars,King1983a}.
For finite-dimensional tensors of $\mathfrak{osp}(m|n)$ it was observed that they follow the same rule:
their superdimension is equal to the dimension of the corresponding 
tensor of $\mathfrak{so}(m-n)$. For example, the superdimension of the skew 3-tensor
of $\mathfrak{osp}(10|4)$ is $6 \cdot 5 \cdot 4 / 6 = 20$, matching the skew 3-tensor of $\mathfrak{so}(6)$.
For a precise statement of when this correspondence holds, see~\cite[Theorem~3.3]{KW}.
Note that also for affine Lie superalgebras this correspondence was examined~\cite{KW2016}.

But the important case of chiral fermions and self dual tensors, which occurs for example in $\mathfrak{osp}(9|1)$ 
supergravity~\cite{GreenSchwarz} remained unsolved. These representations have no simple covariant quantization scheme,
probably because they have no finite-dimensional $\mathfrak{osp}(m|n)$ counter-part.
The aim of the present study is to address this problem.

We show that the chiral spinors and self dual tensors of $\mathfrak{osp}(m|n)$ exist,
that they are infinite-dimensional, and that they also
follow the $\mathfrak{osp}(m|n) \sim \mathfrak{so}(m-n)$
correspondence,
possibly paving the way for a new covariant quantization scheme.

In terms of (the distinguished) Dynkin diagrams of $\mathfrak{osp}(m|n)$, the spinor representation has Dynkin labels 
$[0,0,\ldots,0,1]$ and the self dual tensor $[0,0,\ldots,0,2]$.
In this paper, we shall treat the irreducible representations (irreps) with Dynkin labels $[0,0,\ldots,0,p]$, where $p$ is a positive integer.
Such representations do not appear in Kac's list of finite-dimensional irreps~\cite{Kac1}, 
so they are infinite-dimensional highest weight representations  (see also~\cite{Yamane}).
For infinite-dimensional irreps, one has to be more careful with notions such as dimension and superdimension.
It will be useful to view (super)dimension as some infinite formal power series in a variable $t$, 
such that the coefficient of $t^k$ keeps track of the finite (super)dimension ``at level~$k$'' from the highest weight
(according to some gradation).

The technique used in this paper to compute superdimensions is as follows. 
We consider the branching to the $\mathfrak{gl}(m|n)$ subalgebra (i.e. $\mathfrak{osp}(2m+1|2n) \rightarrow \mathfrak{gl}(m|n)$
and $\mathfrak{osp}(2m|2n) \rightarrow \mathfrak{gl}(m|n)$).
The decomposition of the infinite-dimensional irrep $[0,0,\ldots,0,p]$ according to this branching rule is a (infinite) direct
sum of covariant tensor representations of $\mathfrak{gl}(m|n)$.
For these tensor representations of $\mathfrak{gl}(m|n)$, labeled by a partition $\lambda$, the superdimension is a very simple expression:
it actually reduces to the dimension of a $\mathfrak{gl}(m-n)$ irrep. 
This explains why it is so useful to determine the decomposition with respect to $\mathfrak{gl}(m|n)$.
Note that this is equivalent to expressing the character of the $\mathfrak{osp}(2m(+1)|2n)$ irrep $[0,0,\ldots,0,p]$ 
as an infinite sum of supersymmetric Schur functions
(since the characters of the covariant tensor representations of $\mathfrak{gl}(m|n)$ are given by these Schur functions).

So the main problem is then the determination of the character of the $\mathfrak{osp}(2m(+1)|2n)$ irrep $[0,0,\ldots,0,p]$ in
an appropriate form. 
For this part of the problem, we are lucky.
For $\mathfrak{osp}(2m+1|2n)$, the character of $[0,0,\ldots,0,p]$ has been determined in~\cite{parast}.
In fact, the irrep $[0,0,\ldots,0,p]$ is (equivalent to) the ``irreducible parastatistics Fock space of order~$p$''.
A combined system of $m$ parafermions and $n$ parabosons generate the algebra $\mathfrak{osp}(2m+1|2n)$,
and the Fock space $V(p)$ of order $p$ (constructed in~\cite{parast}) corresponds exactly to the irrep $[0,0,\ldots,0,p]$.
For the case of $\mathfrak{osp}(2m|2n)$, the character of $[0,0,\ldots,0,p]$ can be obtained by 
considering the decomposition $\mathfrak{osp}(2m+1|2n)\supset\mathfrak{osp}(2m|2n)$.

The structure of the paper is as follows.
In section~\ref{sec-example} we describe, as an introductory example, the construction of $\mathfrak{osp}(m|2n)$ spinors.
The ideas of this construction appeared already in~\cite{TM}, but are reviewed here because they open the way to the more general problem discussed in this paper.
In the following sections, we describe some of the mathematical notions needed here.
In section~\ref{sec2}, we recall the basic notation and terminology related to partitions, 
remind the reader of symmetric and supersymmetric Schur functions and their relation to $\mathfrak{gl}(n)$ and $\mathfrak{gl}(m|n)$ irreps,
and recall some useful (super)dimension formulas for the $\mathfrak{gl}$ case.
Section~\ref{sec3} deals with (super)dimensions of some infinite-dimensional representations of $\mathfrak{osp}(1|2n)$, 
which is a special case among the orthosymplectic Lie superalgebras.
We shall see that the technique and results of this section will be crucial for the remainder of the paper.
The core of the paper is in sections~\ref{sec4} ($\mathfrak{osp}(2m+1|2n)$) and~\ref{sec5} ($\mathfrak{osp}(2m|2n)$).
The main results are as follows:
the superdimension of the $\mathfrak{osp}(2m+1|2n)$ irrep $[0,0,\ldots,0,p]$ is equal to the dimension of the 
$\mathfrak{so}(2m+1-2n)$ irrep $[0,0,\ldots,0,p]$. Herein, if $2m+1-2n$ is negative, the algebra $\mathfrak{so}(2m+1-2n)$
should be interpreted as $\mathfrak{osp}(1|2n-2m)$.
Secondly, the superdimension of the $\mathfrak{osp}(2m|2n)$ irrep $[0,0,\ldots,0,p]$ is equal to the dimension of the 
$\mathfrak{so}(2m-2n)$ irrep $[0,0,\ldots,0,p]$ (or $[0,0,\ldots,p,0]$). 
In this case, if $2m-2n$ is negative, the algebra $\mathfrak{so}(2m-2n)$
should be interpreted as $\mathfrak{sp}(2n-2m)$.
At the end of the paper, we also include a section on $D(2,1;\alpha)$, in order to cover all type~II orthosymplectic Lie superalgebras,
and we give a summary of the main results in terms of Dynkin diagrams.

\section{Construction of $\mathfrak{osp}(m|2n)$ spinors}
\label{sec-example}

In his famous lectures on the theory of spinors~\cite{Cartan}, Cartan has shown that there is no finite-dimensional spinor in curved space. 
But they exist as infinite-dimensional representations, analysed much later by Ne'eman~\cite{Ne'eman}.
In the same way, Kac~\cite{Kac1} has shown that the orthosymplectic Lie superalgebras have no finite-dimensional spinors.
Their infinite-dimensional counterpart, however, can be constructed very simply as follows~\cite{TM}.

Consider the vector space $P^n$ of polynomials in $n$ formal variables $x^1,\ldots,x^n$ with complex coefficients. 
This space is naturally graded by the total degree in the variables
\begin{equation}
   P^n = \bigoplus_{k=0}^\infty P_k,
\label{cliff-w1}
\end{equation}
where each subspace $P_k$ has dimension $\binom{n+k-1}{k}$.
Its partition function, or $t$-dimension (see later for a formal definition), is
\begin{equation}
  \dim_t (P^n) = \sum_{k=0}^\infty \dim (P_k) t^k = \frac {1}{ (1 - t)^n}.
\label{cliff-w2}
\end{equation}

The $2n$ operators of multiplication and partial derivation relative to $x^i$
\begin{equation}
   \gamma_{2i - 1} = x^i , \qquad
 \gamma_{2i} = 2 \frac {\partial}{\partial {x^i}} ,\qquad i = 1,2,... n
\label{cliff-w3}
\end{equation}
form a Weyl algebra   
\begin{equation}
   \gamma_i \gamma_j - \gamma_j \gamma_i = 2 \epsilon_{ij} I
\label{cliff-w4}
\end{equation}
where $I$ is the identity and $\epsilon$ is a $2n$ dimensional block diagonal matrix
where each $2 \times 2$ diagonal block is of the form 
\[ 
\left[ \begin{array}{cc}0 & -1 \\ 1 & 0 \end{array} \right].
\]
In other words, $\epsilon$ is the canonical skew symmetric metric fixed by the $\mathfrak{sp}(2n)$
symplectic Lie algebra. And indeed, the $2n (2n+1)/2$ symmetric bilinear operators
\begin{equation}
  J_{ij} = \frac 1 4  (\gamma_i \gamma_j + \gamma_j \gamma_i) \qquad i,j = 1,2,... 2n
\label{cliff-w5}
\end{equation}
satisfy the $\mathfrak{sp}(2n)$ commutation rules
\begin{equation}
  [J_{ij}, J_{kl}] = J_{ij} J_{kl} - J_{kl} J_{ij} = 
  \epsilon_{jk} J_{il}
  + \epsilon_{jl} J_{ik}
  + \epsilon_{il} J_{jk}
  + \epsilon_{ik} J_{jl}.
\label{cliff-w6}
\end{equation}
The Chevalley generators  $e_i, f_i, h_i, \; i=1,2....n $ of $\mathfrak{sp}(2n)$, 
satisfying the standard Chevalley-Serre relations~\cite{Serre},
are represented by the operators
\begin{align*}
& e_i = x^{i+1} \frac{\partial}{\partial x^i},  \;\;  f_i = x^{i} \frac{\partial}{\partial x^{i+1}}, \;\;
h_i = x^{i+1} \frac{\partial}{\partial x^{i+1}} - x^{i} \frac{\partial}{\partial x^{i}} , \;\;  i=1,\ldots ,n-1\\
& e_n = \frac{1}{2}\left(\frac{\partial}{\partial x^{n}} \right)^2, \;\; f_n = - \frac 1 2 (x^n)^2, \;\;
h_n = - \frac 1 2 (x^n \frac{\partial}{\partial x^{n}}  + \frac{\partial}{\partial x^{n}} x^n).
\end{align*}
Since  these operators map polynomials of even (resp. odd) degree in $x$ into polynomials of even (resp. odd) degree in $x$
one obtains the two irreducible so-called
metaplectic representations characterized by their highest weight $1$ and $x_n$, the only monomial
belonging to the kernel of all the $e_i$. By inspection, the Dynkin weights are:
\begin{align*}
& \raisebox{-5mm}{\includegraphics{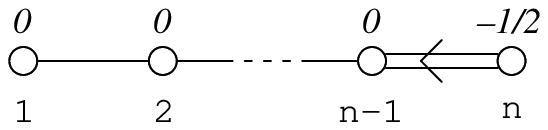}} \\
& \\
& \raisebox{-5mm}{\includegraphics{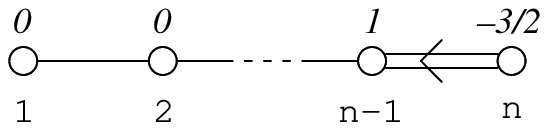}}\\
\end{align*}
In a similar way, the orthogonal Lie algebra $\mathfrak{so}(2m)$ can be realized by means of operators acting on polynomials of
anti-commuting variables. For this purpose, consider the vector space $Q^m$ of polynomials in $m$ formal anti-commuting variables
$\theta^{\alpha}$ ($\alpha=1,2,\ldots,m$) with complex coefficients.
Since $(\theta^\alpha)^2 = 0$, this space is finite-dimensional
\begin{equation}
   \dim (Q ^m) = 2^m.
\label{cliff-c1}
\end{equation}
The following $2m$ operators 
\begin{align*}
 &  \gamma_{2\alpha - 1} = \theta^{\alpha} +  \frac \partial {\partial {\theta^{\alpha}}}, \\
 &  \gamma_{2\alpha} = i (\theta^{\alpha} -  \frac \partial {\partial {\theta^{\alpha}}}), \qquad \alpha = 1,2,... m
 \label{cliff-c2}
\end{align*}
represent the $2m$ dimensional Clifford algebra
\begin{equation}
   \gamma_{\mu} \gamma_{\nu} + \gamma_{\nu} \gamma_{\mu} = 2 g_{\mu \nu} I,         \qquad  \mu, \nu = 1,2.... 2m
 \label{cliff-c3}
\end{equation}
where $I$ is the identity and $g_{\mu \nu}$ is a $2m$ dimensional identity matrix.
In other words, $g_{\mu \nu}$ is the canonical symmetric metric fixed by the $\mathfrak{so}(2m)$
orthogonal Lie algebra. And indeed, the $2m (2m-1)/2$ antisymmetric bilinear operators
\begin{equation}
  J_{\mu \nu} = \frac 1 4  (\gamma_{\mu} \gamma_{\nu} - \gamma_{\nu} \gamma_{\mu}) , \qquad   \mu, \nu = 1,2.... 2m
 \label{cliff-c4}
\end{equation}
satisfy the $\mathfrak{so}(2m)$ commutation rules
\begin{equation}
  [J_{\mu \nu}, J_{\rho \sigma}] = J_{\mu \nu} J_{\rho \sigma} - J_{\rho \sigma} J_{\mu \nu}= 
  g_{\nu \rho} J_{\mu \sigma}
  - g_{\nu \sigma} J_{\mu \rho}
  + g_{\mu \sigma} J_{\nu \rho}
  - g_{\mu \rho} J_{\nu \sigma}.
 \label{cliff-c5}
\end{equation}
In other words, using polynomials in anti-commuting variables, one obtains a simple construction 
of the well known $2^m$-dimensional Dirac matrices in $2m$ dimensions.

Since the $J_{\mu \nu}$ operators are even in $\theta$, the representation space $Q^m$ splits into the
even and odd polynomials, which correspond to the left and right spinors,
each of dimension $2^{m-1}$. The $\mathfrak{so}(2m+1)$ algebra is obtained by defining the operator $\Gamma_5$ and adding
the operators $J_{\mu 5}=-J_{5\mu}$:
\begin{equation}
\Gamma_5 = \gamma_1\gamma_2\cdots\gamma_{2m}, \\
\qquad  J_{\mu 5}= \gamma_{\mu}  \Gamma_5= -\Gamma_5 \gamma_{\mu} = -J_{5\mu}  , \\
\qquad \Gamma_5^2 = (-1)^{m}.
\label{cliff-c5a}
\end{equation}
Since $\Gamma_5$ is even in the $\theta$'s, the $J_{\mu 5}$ operators are odd 
and they connect the even and odd polynomials, so the space $Q^m$ is irreducible and the
dimension of the spinors of $\mathfrak{so}(2m+1)$ is $2^m$.

The parallel between the $\mathfrak{so}$ and $\mathfrak{sp}$ construction is striking, but notice that
the statistics is counter-intuitive: in spinor space $\mathfrak{sp}(2n)$ acts on ``bosonic polynomials''
and $\mathfrak{so}(2m+1)$ on Grassmann or ``fermionic polynomials''.

The generalization to the orthosymplectic Lie superalgebra $\mathfrak{osp}(2m|2n)$ is now immediate.
One considers the space $P^{m|n}$ of superpolynomials in $m$ anti-commuting and $n$ commuting variables.
As we will see later in general, this corresponds to the spinor representation of $\mathfrak{osp}(2m|2n)$. 
Using the same gradation by the total degree in $x$, the $t$-dimension is
\begin{equation}
    \dim_t (P^{m|n}) = \frac {2^m}{ (1 - t) ^ n}.
\label{cliff-c6}
\end{equation}
The superdimension (difference between dimension of even and odd subspaces)  is
\begin{equation}
    \sdim (P^{m|n}) = \dim_{-1} (P^{m|n}) = 2 ^{m - n} .
 \label{cliff-c7}
\end{equation}
So for $m\geq n$, the superdimension of the (infinite-dimensional) spinor representation of $\mathfrak{osp}(2m|2n)$
coincides with the (finite) dimension~\eqref{cliff-c1} of the spinor representation of $\mathfrak{so}(2m-2n)$,
fulfilling the correspondence being studied in this paper.

To construct the $\mathfrak{osp}(2m|2n)$ generators, one considers the  $m(2m-1)$ antisymmetric $J_{\mu \nu}$, the  $n(2n+1)$ symmetric $J_{ij}$ and adds the $4mn$ odd operators
\begin{equation}
    J_{\mu i} = \gamma_{\mu} \gamma_i , \qquad  \mu = 1,2.... 2m , \qquad i = 1,2,... 2n.¯
 \label{cliff-wg1}
\end{equation}
Their anticommutators close on the previous even operators, completing the construction of the $\mathfrak{osp}(2m|2n)$ superspinor representation:
\begin{equation}
    \{ J_{\mu i}, J_{\nu j} \} = J_{\mu i} J_{\nu j} + J_{\nu j}  J_{\mu i}
    = 4 \epsilon_{ij}  J_{\mu \nu} + 4 g_{\mu \nu} J_{ij}.
 \label{cliff-wg2}
\end{equation}
The superspinor representation space $P^{m|n}$ naturally splits in four components
\begin{equation}
P^{m|n} = P^{m|n}_{0,0} \oplus P^{m|n}_{1,0} \oplus P^{m|n}_{0,1} \oplus P^{m|n}_{1,1},
\end{equation}
where $P^{m|n}_{i,j}$ denotes the subspace of polynomials of total even (resp.\ odd) degree in the variables $\theta$ for $i=0$ (resp.\ $i=1$)
and of total even (resp.\ odd) degree in the variables $x$ for $j=0$ (resp.\ $j=1$).
Since the  odd operators $J_{\mu i}$ are of odd degree in $x$ and $\theta$, they map these subspaces onto each other.
On the other hand, these operators are of total even degree (in $x$ and $\theta$ combined),
so under the action of $\mathfrak{osp}(2m|2n)$ the superspinor representation space $P^{m|n}$ still splits into a ``left sector''
$P^{m|n}_{0,0} \oplus P^{m|n}_{1,1}$ and a ``right sector'' $P^{m|n}_{1,0} \oplus P^{m|n}_{0,1}$, each of superdimension $2^{m - n - 1}$.

The case $\mathfrak{osp}(2m+1|2n)$ follows by adding the $\mathfrak{so}(2m+1)$ operator $\Gamma_5$ as described above,
and in this case the space $P^{m|n}$ is irreducible of superdimension $2^{m-n}$.

In the following sections we shall be dealing with the self dual tensors of the orthosymplectic Lie superalgebras and their generalizations.
These can be considered as the top irreducible component of the $p^{th}$ symmetric tensor product of the superspinor.
These self dual tensors are also infinite-dimensional and follow the $\mathfrak{osp}(m|n) \sim \mathfrak{so}(m-n)$ correspondence.
Notice that we have given in this section a construction of the superspinor, with the explicit action of all the generators of $\mathfrak{osp}(m|n)$.
As far as we know, there is no such simple construction for the self dual tensors.
In the remainder of the paper, we only give the characters of the irreducible self dual tensors, and
that is already sufficient to establish the correspondence.

\section{Preliminaries}
\label{sec2}
\subsection{Partitions, symmetric functions and supersymmetric functions}

We need in this paper some basic notions on partitions and symmetric functions, see~\cite{Mac} as a standard reference.
A partition $\lambda=(\lambda_1,\lambda_2,\ldots,\lambda_n)$ of weight $|\lambda|$ and length $\ell(\lambda)\leq n$
is a sequence of non-negative integers satisfying the condition $\lambda_1\geq\lambda_2\geq\cdots\geq\lambda_n$, such that their
sum is $|\lambda|$, and $\lambda_i>0$ if and only if $i\leq \ell(\lambda)$. 
To each such partition there corresponds a Young diagram $F^\lambda$ consisting of $|\lambda|$ boxes arranged in $\ell(\lambda)$ left-adjusted
rows of lengths $\lambda_i$ for $i=1,2,\ldots,\ell(\lambda)$. 
For example, the Young diagram of $\lambda=(5,4,4,2)$ is given by
\[
\begin{Young}
&&&&\cr
&&&\cr
&&&\cr
&\cr
\end{Young}
\]
The conjugate partition $\lambda'$ corresponds to the Young diagram of $\lambda$ reflected about the main diagonal. 
In other words, $\lambda_j'$ is the length of column~$j$ of $F^\lambda$. For the above example, $\lambda'=(4,4,3,3,1)$.

An important notion is the Frobenius notation~\cite{Mac} of a partition $\lambda$. If $F^\lambda$ has $r=r(\lambda)$ boxes on the diagonal, 
$r$ is said to be the rank of $\lambda$. 
In the above example, $r=3$, denoted by crosses in the diagonal boxes:
\[
\begin{Young}
$\times$&&&&\cr
&$\times$&&\cr
&&$\times$&\cr
&\cr
\end{Young}
\]
The arm lengths $a_k=\lambda_k-k$ and leg lengths $b_k=\lambda_k'-k$ ($k=1,\ldots,r$) refer to the remaining boxes to the right or below the $k$th diagonal box, where $a_1>a_2>\cdots>a_r\geq 0$ and $b_1>b_2>\cdots>b_r\geq 0$.
The Frobenius notation of $\lambda$ is then
\[
\lambda=\left( \myatop{a_1\ a_2 \cdots a_r}{b_1\ b_2 \cdots b_r} \right).
\]
Note that for our example we have $\lambda = \left( \myatop{4\, 2\, 1}{3\, 2\, 0} \right)$.

Denote by $\Lambda_n$ the ring of symmetric polynomials with integer coefficients in the $n$ independent variables $x=(x_1,x_2,\ldots,x_n)$.
The Schur functions~\cite{Mac} or $S$-functions $s_\lambda(x)$ form a $\mathbb{Z}$-basis of $\Lambda_n$, where $\lambda$ runs over the set of all partitions of length at most $n$.
There are various ways to define Schur functions. For a partition $\lambda$ with $\ell(\lambda)\leq n$, one has
\begin{equation}
s_\lambda(x) = \frac{\det (x_i^{\lambda_j+n-j})_{1\leq i,j\leq n}}{\det (x_i^{n-j})_{1\leq i,j\leq n}}.
\label{schur}
\end{equation}
If $\ell(\lambda)>n$, one puts $s_\lambda(x)=0$. Clearly, for the zero partition $\lambda=(0)$ (which is the only partition with 
Frobenius rank $r=0$) one has $s_{(0)}(x)=1$.

In terms of two sets of variables $x=(x_1,\ldots,x_m)$ and $y=(y_1,\ldots,y_n)$, one can define the ring $\Lambda_{m,n}$ of supersymmetric polynomials with integer coefficients~\cite{Berele}. This ring consists of all double symmetric polynomials in $x$ and $y$ (elements of $\Lambda_m\otimes \Lambda_n$) that satisfy the so-called cancellation property (i.e.\ when the substitution $x_1=t$, $y_1=-t$ is made in an element $p$ of $\Lambda_m\otimes \Lambda_n$, the resulting polynomial is independent of $t$). 
For a partition $\lambda$, one can define supersymmetric Schur functions $s_\lambda(x|y)$ belonging to $\Lambda_{m,n}$.
One possible definition is~\cite{Berele,King1983}
\[
s_\lambda(x|y)= \sum_{\mu,\nu} c^{\lambda}_{\mu,\nu} s_\mu(x) s_{\nu'}(y),
\]
where the sum is over all partitions $\mu$ ($\ell(\mu)\leq m$) and $\nu$ ($\ell(\nu')\leq n$), and $c^{\lambda}_{\mu,\nu}$ are the Littlewood-Richardson coefficients~\cite{Mac}.
The polynomials $s_\lambda(x|y)$ are identically zero when $\lambda_{m+1}>n$. Denote by ${\cal H}_{m,n}$ the set of all partitions with $\lambda_{m+1}\leq n$, i.e.\ the partitions (with their Young diagram) inside the $(m,n)$-hook. The set of $s_\lambda(x|y)$ with $\lambda\in{\cal H}_{m,n}$ form a $\mathbb{Z}$-basis of $\Lambda_{m,n}$.

\subsection{Dimension formulas for $\mathfrak{gl}(n)$ and superdimensions for $\mathfrak{gl}(m|n)$}

A finite-dimensional irreducible representation of the Lie algebra $\mathfrak{gl}(n)$ is characterized 
by a partition $\lambda$ with $\ell(\lambda)\leq n$. 
In terms of the standard basis $\epsilon_1,\ldots,\epsilon_n$ of the weight space of $\mathfrak{gl}(n)$, the highest weight of this
representation is $\sum_{i=1}^n \lambda_i \epsilon_i$, and the representation space will be denoted by $V_{\mathfrak{gl}(n)}^{\lambda}$. 
Weyl's character formula for such representations corresponds to the right hand side of~\eqref{schur}, and thus
\begin{equation}
\ch V_{\mathfrak{gl}(n)}^{\lambda}  = s_\lambda(x),
\end{equation}
where $x_i=\hbox{e}^{\epsilon_i}$.

There are two useful formulas for the dimension of $V_{\mathfrak{gl}(n)}^{\lambda}$, by specializing all $x_i=1$ in the Schur function $s_\lambda$.
One is Weyl's dimension formula:
\begin{equation}
\dim V_{\mathfrak{gl}(n)}^{\lambda} = \prod_{1\leq i<j\leq n} \frac{(\lambda_i-\lambda_j+j-i)}{(j-i)}.
\label{dimWeyl}
\end{equation}
The other expression uses hook-lengths:
\begin{equation}
\dim V_{\mathfrak{gl}(n)}^{\lambda} = \prod_{(i,j)\in\lambda} \frac{n+j-i}{h_{ij}}= 
\prod_{(i,j)\in\lambda} \frac{n+j-i}{\lambda_i+\lambda_j'-i-j+1}.
\label{dimhook}
\end{equation}
Herein, the product runs over all boxes $(i,j)$ of the Young diagram $F^\lambda$, and $h_{ij}=\lambda_i+\lambda_j'-i-j+1$ is the number 
of boxes of the hook with hook box $(i,j)$. 
For example, for the partition $(5,4,4,2)$ and $n=5$, the numbers $(n+j-i)$ and $h_{ij}$ read, respectively,
\[
\begin{Young}
5&6&7&8&9\cr
4&5&6&7\cr
3&4&5&6\cr
2&3\cr
\end{Young} \qquad \qquad
\begin{Young}
8&7&5&4&1\cr
6&5&3&2\cr
5&4&2&1\cr
2&1\cr
\end{Young}
\]
Note, if $\lambda$ is a partition with $\ell(\lambda)>n$, \eqref{dimhook} gives automatically 0, whereas in~\eqref{dimWeyl} one has to impose
that $\dim V_{\mathfrak{gl}(n)}^{\lambda} = 0$.

Just as the functions $s_\lambda(x)$ are characters of irreducible representations (or simple modules) of the Lie algebra $\mathfrak{gl}(n)$,
the supersymmetric Schur functions are characters of a class of simple modules of the Lie superalgebra $\mathfrak{gl}(m|n)$,
namely of the covariant representations~\cite{Berele}.
For a partition $\lambda\in{\cal H}_{m,n}$, the corresponding covariant representation will be denoted by $V_{\mathfrak{gl}(m|n)}^{\lambda}$.
In terms of the standard basis $\epsilon_1,\ldots,\epsilon_m,\ \delta_1,\ldots,\delta_n$ of the weight space of $\mathfrak{gl}(m|n)$, 
the highest weight of this representation is $\sum_{i=1}^m \lambda_i \epsilon_i + \sum_{j=1}^n \max(\lambda_j'-m,0)\delta_j$.
The main result of~\cite{Berele} is
\begin{equation}
\ch V_{\mathfrak{gl}(m|n)}^{\lambda} = s_\lambda(x|y),
\end{equation}
where $x_i=\hbox{e}^{\epsilon_i}$ and $y_j=\hbox{e}^{\delta_j}$.

Any Lie superalgebra $\mathfrak{g}$ is $\mathbb{Z}_2$-graded: $\mathfrak{g}=\mathfrak{g}_{\bar 0}\oplus\mathfrak{g}_{\bar 1}$.
A Lie superalgebra module or representation $V$ is also $\mathbb{Z}_2$-graded: $V=V_{\bar 0}\oplus V_{\bar 1}$. 
In our convention, the highest weight vector $v$ of $V$ will always be an even vector ($v\in V_{\bar 0}$).
When $V$ is finite-dimensional, one can speak of the dimension and superdimension of $V$:
\[
\dim V = \dim V_{\bar 0}+ \dim V_{\bar 1},\qquad \sdim V = \dim V_{\bar 0} - \dim V_{\bar 1}.
\]
Dimension formulas for covariant representations of $\mathfrak{gl}(m|n)$ are not particularly simple~\cite{Moens}, but there is a great simplification when it comes to superdimensions~\cite{King1983}. The result depends on whether $m$ is greater than, equal to, or less than $n$. 
It can be summarized as follows:
\begin{equation}
\sdim V_{\mathfrak{gl}(n+k|n)}^{\lambda} =\dim V_{\mathfrak{gl}(k)}^{\lambda}, \qquad
\sdim V_{\mathfrak{gl}(m|m+k)}^{\lambda} =(-1)^{|\lambda|}\dim V_{\mathfrak{gl}(k)}^{\lambda'}.
\label{dim-sdim}
\end{equation}
In particular, when $m=n$, $\sdim V_{\mathfrak{gl}(n|n)}^{\lambda} =0$ unless $\lambda$ is the zero partition $(0)$ (then 
$V_{\mathfrak{gl}(n|n)}^{(0)}$ is the trivial module with $\sdim V_{\mathfrak{gl}(n|n)}^{(0)} =1$).
Note that~\eqref{dim-sdim} implies: when $\ell(\lambda)>k$ then $\sdim V_{\mathfrak{gl}(n+k|n)}^{\lambda}=0$;
when $\lambda_1>k$ then $\sdim V_{\mathfrak{gl}(m|m+k)}^{\lambda}=0$.

\section{The Lie superalgebra $\mathfrak{osp}(1|2n)$}
\label{sec3}

Although the main aim is to study superdimensions of certain infinite-dimensional representations of the Lie superalgebras 
$B(m,n)=\mathfrak{osp}(2m+1|2n)$ and $D(m,n)=\mathfrak{osp}(2m|2n)$, it will be useful to start with the special case $B(0,n)=\mathfrak{osp}(1|2n)$.
Let us first fix some notation~\cite{Kac,Kac1,dict}. 
In the common basis $\delta_j$ for the weight space of $\mathfrak{osp}(1|2n)$, the odd roots are given by $\pm\delta_j$ ($j=1,\ldots,n$), and the even roots are $\delta_i-\delta_j$ ($i\ne j$) and $\pm(\delta_i+\delta_j)$. 
The simple roots are 
\begin{equation}
\delta_1-\delta_2,\  \delta_2-\delta_3, \ldots, \delta_{n-1}-\delta_n,\ \delta_n. 
\label{osp12n}
\end{equation}
It will be helpful to use the following $\mathbb{Z}$-gradation of $\mathfrak{g}= \mathfrak{osp}(1|2n)$: 
$\mathfrak{g}=\mathfrak{g}_{-2}\oplus\mathfrak{g}_{-1}\oplus\mathfrak{g}_0\oplus\mathfrak{g}_{+1}\oplus\mathfrak{g}_{+2}$, where each
$\mathfrak{g}_j$ is spanned by the root vectors corresponding to the following roots:
\[
\begin{array}{ccccc}
\phantom{mm}\mathfrak{g}_{-2}\phantom{mm} &\phantom{mm}\mathfrak{g}_{-1}\phantom{mm} &\phantom{mm}\mathfrak{g}_{0}\phantom{mm} &\phantom{mm}\mathfrak{g}_{+1}\phantom{mm} &\phantom{mm}\mathfrak{g}_{+2}\phantom{mm} \\
-\delta_i-\delta_j & -\delta_i & \delta_i-\delta_j & \delta_i & \delta_i+\delta_j
\end{array}
\]
Note that $\mathfrak{g}_{0}=\mathfrak{gl}(n)$. 

We will consider some infinite-dimensional highest weight representations $V$ of $\mathfrak{g}$, so at first sight it might be strange to speak of the dimension and superdimension of $V$. However, if the action of $\mathfrak{g}_0=\mathfrak{gl}(n)$ on the highest weight vector $v$ of $V$ corresponds to a finite-dimensional $\mathfrak{g}_0$ module $V_0$, then the $\mathbb{Z}$-gradation of $\mathfrak{g}$ induces a $\mathbb{Z}$-gradation of $V$:
\[
V=V_0\oplus V_{-1} \oplus V_{-2} \oplus \cdots
\]
in terms of finite-dimensional $\mathfrak{g}_0$ modules. 
Then one defines:
\begin{equation}
\dim_t (V) = \sum_{i=0}^\infty \dim V_{-i}\ t^i, \qquad \sdim_t (V) = \sum_{i=0}^\infty \dim V_{-i}\ (-t)^i.
\end{equation}
Clearly $\sdim_t (V) =\dim_{-t} (V)$, and $\dim_t (V)$ is a formal power series in the variable $t$ such that the coefficient of $t^k$ counts the dimension ``at level $k$'' according to the $\mathbb{Z}$-gradation.

For reasons that will become clear, we will consider the irreducible highest weight representation with highest weight given by
$(-\frac{p}{2},-\frac{p}{2},\ldots, -\frac{p}{2})= \sum_{i=1}^n (-\frac{p}{2})\delta_i$, where $p$ is a non-negative integer. 
This representation will be denoted by $V=V_{\mathfrak{osp}(1|2n)}^{(-p/2)^n}$.
Its Dynkin labels according to the simple roots system~\eqref{osp12n} are $[0,0,\ldots,0,-p]$.
Clearly, with this choice of highest weight one has $\dim V_0=1$.
Obviously, the structure of the irreducible highest weight representation with highest weight $(-\frac{p}{2},-\frac{p}{2},\ldots, -\frac{p}{2})$ is completely the same as the irreducible lowest weight representation with lowest weight $(\frac{p}{2},\frac{p}{2},\ldots, \frac{p}{2})$.
This last representation is well known: it is the paraboson representation of order $p$.
Its structure and character have been determined in~\cite{paraboson}. 
Using the notation $x_i=\hbox{e}^{-\delta_i}$, the following character formula holds:
\begin{equation}
\ch V_{\mathfrak{osp}(1|2n)}^{(-p/2)^n} = (x_1\cdots x_n)^{p/2} \sum_{\lambda,\ \ell(\lambda)\leq p} s_\lambda(x).
\label{char-osp-1}
\end{equation}
This is an infinite sum over all partitions of length at most $p$. Since $s_\lambda(x)=0$ if $\ell(\lambda)>n$, the sum is actually over all partitions satisfying $\ell(\lambda)\leq\min(n,p)$.

In order to give a first expression of $\dim_t V_{\mathfrak{osp}(1|2n)}^{(-p/2)^n}$ one should (apart from the factor $(x_1\cdots x_n)^{p/2}$) specify $x_i=t$ in~\eqref{char-osp-1}. Thus one finds:
\begin{equation}
\dim_t V_{\mathfrak{osp}(1|2n)}^{(-p/2)^n} = \sum_{\lambda,\ \ell(\lambda)\leq\min(n,p)} \dim V_{\mathfrak{gl}(n)}^{\lambda} t^{|\lambda|}.
\label{tdim-osp1}
\end{equation}
This is an infinite sum; it can be simplified using the alternative character formula also determined in~\cite{paraboson}:
\begin{equation}
\ch V_{\mathfrak{osp}(1|2n)}^{(-p/2)^n} = (x_1\cdots x_n)^{p/2} \frac{E_{n,p}(x)}{\prod_i(1-x_i) \prod_{i<j}(1-x_ix_j)}.
\label{char-osp-2}
\end{equation}
Herein, 
\[
E_{n,p}(x) = \sum_\eta (-1)^{c_\eta} s_\eta(x),
\]
where the sum is over all partitions $\eta$ of Frobenius form
\[
\eta=\left( \myatop{a_1}{a_1+p}\ \myatop{a_2}{a_2+p} \cdots \myatop{a_r}{a_r+p} \right)
\]
and $c_\eta=a_1+\cdots+a_r+r=|a|+r$. So the sum in $E_{n,p}(x)$ is over all self-conjugate partitions to which $r$ ``legs'' of $p$ boxes are added.
Note that $E_{n,p}(x)=1$ if $p\geq n$. In the following, this trivial case will be left out and we shall deal only with $p<n$.

Specifying $x_i=t$ in \eqref{char-osp-2} now yields
\begin{equation}
\dim_t V_{\mathfrak{osp}(1|2n)}^{(-p/2)^n} = \frac{E_{n,p}(t,\ldots,t)}{(1-t)^n (1-t^2)^{n(n-1)/2}}.
\label{tdim-osp2}
\end{equation}
So, let us therefore consider $E_{n,p}(x)$ in more detail, starting with $E_{n,0}(x)$. Since $s_\lambda(x)=0$ when $\ell(\lambda)>n$, 
the expression $E_{n,0}(x)$ has only a finite number of terms:
\begin{equation}
E_{n,0}(x) = \sum_{r=0}^n\  \sum_{n-1\geq a_1>a_2>\cdots>a_r\geq 0} (-1)^{|a|+r} 
s_{\left(\myatop{a_1}{a_1}\, \myatop{a_2}{a_2} \cdots \myatop{a_r}{a_r} \right)}(x).
\end{equation}
It is easy to see that this expression has $2^n$ terms. 
Similarly, $E_{n,p}(x)$ has $2^{n-p}$ terms, and
\begin{equation}
E_{n,p}(x) = \sum_{r=0}^{n-p}\  \sum_{n-p-1\geq a_1>a_2>\cdots>a_r\geq 0} (-1)^{|a|+r} 
s_{\left(\myatop{a_1}{a_1+p}\, \myatop{a_2}{a_2+p} \cdots \myatop{a_r}{a_r+p} \right)}(x).
\end{equation}
Let us now turn to the dimension formula. In general, for a partition $\lambda$ given in Frobenius notation
\[
\lambda=\left( \myatop{a_1\ a_2 \cdots a_r}{b_1\ b_2 \cdots b_r} \right) \equiv \left( \myatop{a}{b} \right)
\]
one can write
\[
s_\lambda(t,\ldots,t) = t^{r+|a|+|b|} s_\lambda(1,\ldots,1) = t^{r+|a|+|b|} \dim V_{\mathfrak{gl}(n)}^{\lambda}.
\]
Using~\eqref{dimhook}, one finds
\begin{equation}
\dim V_{\mathfrak{gl}(n)}^{\lambda}= \dim V_{\mathfrak{gl}(n)}^{\left( \myatop{a}{b} \right)}= 
{\prod_{i=1}^r \frac{(n+a_i)!}{(n-b_i-1)!}} {\Biggl /}
{\frac{\prod_{i=1}^r a_i!b_i! \prod_{i,j=1}^r (a_i+b_j+1)}{\prod_{1\leq i<j\leq r} (a_i-a_j)(b_i-b_j)}}.
\label{dim-F-gl}
\end{equation}
And hence
\begin{equation}
E_{n,p}(t,\ldots,t)= \sum_{r=0}^{n-p}  \sum_{n-p-1\geq a_1>a_2>\cdots>a_r\geq 0} (-1)^{|a|+r} 
 \dim V_{\mathfrak{gl}(n)}^{\left( \myatop{a}{a+p} \right)} t^{2|a|+(p+1)r}.
\label{Enpt}
\end{equation}
Let us complete this section by means of some examples:
\begin{align*}
E_{1,0}(t,\ldots,t) &= 1-t\\
E_{2,0}(t,\ldots,t) &= 1-2t+2t^3-t^4=(1-t)^2(1-t^2)\\
E_{3,0}(t,\ldots,t) &= 1-3t+8t^3-6t^4-6t^5+8t^6-3t^8+t^9=(1-t)^3(1-t^2)^3\\
E_{3,1}(t,\ldots,t) &= 1-3t^2+3t^4-t^6=(1-t^2)^3\\
E_{3,2}(t,\ldots,t) &= 1-t^3\\
E_{4,1}(t,\ldots,t) &= 1-6t^2+15t^4-20t^6+15t^8-6t^{10}+t^{12}=(1-t^2)^6
\end{align*}
Of course, due to the general factorization of $E_{n,0}(x)$~\cite{King1990}, one has $E_{n,0}(t,\ldots,t)=(1-t)^2(1-t^2)^{n(n-1)/2}$.
Some other general expressions following from~\eqref{Enpt} are:
\begin{align*}
E_{n,1}(t,\ldots,t) &= (1-t^2)^{n(n-1)/2}\\
E_{p+1,p}(t,\ldots,t) &= 1-t^{p+1}\\
E_{p+2,p}(t,\ldots,t) &= 1-(p+2)t^{p+1}+(p+2)t^{p+3}-t^{2p+4}\\
E_{p+3,p}(t,\ldots,t) &= 1-\frac{(p+2)(p+3)}{2}t^{p+1}+(p+2)(p+4)t^{p+3}-\frac{(p+3)(p+4)}{2}t^{p+5}\\
& -\frac{(p+3)(p+4)}{2}t^{2p+4}+(p+2)(p+4)t^{2p+6}-\frac{(p+2)(p+3)}{2}t^{2p+8}-t^{3p+9}
\end{align*}
These were a few examples of $E_{n,p}(t,\ldots,t)$. Including denominators according to~\eqref{tdim-osp2} 
one finds the actual dimensions, e.g.:
\begin{align*}
\dim_t V_{\mathfrak{osp}(1|6)}^{(0,0,0)} &= 1\\
\dim_t V_{\mathfrak{osp}(1|6)}^{(-1/2,-1/2,-1/2)}  &= \frac{1-3t^2+3t^4-t^6}{(1-t)^3(1-t^2)^3}=\frac{1}{(1-t)^3}
= 1+3t+6t^2+10t^3+15t^4+\cdots\\
\dim_t V_{\mathfrak{osp}(1|6)}^{(-1,-1,-1)}  &= \frac{1-t^3}{(1-t)^3(1-t^2)^3}
= 1+3t+9t^2+18t^3+36t^4+\cdots\\
\dim_t V_{\mathfrak{osp}(1|6)}^{(-3/2,-3/2,-3/2)}  &= \frac{1}{(1-t)^3(1-t^2)^3}= 1+3t+9t^2+19t^3+39t^4+\cdots
\end{align*}

\section{The Lie superalgebra $\mathfrak{osp}(2m+1|2n)$}
\label{sec4}

We will now treat one of the two main cases, the Lie superalgebra $B(m,n)=\mathfrak{osp}(2m+1|2n)$.
The weight space of $\mathfrak{osp}(2m+1|2n)$ has basis $\epsilon_1,\ldots,\epsilon_m,\delta_1,\ldots,\delta_n$, and the distinguished set
of simple roots is~\cite{Kac,dict} 
\begin{equation}
\delta_1-\delta_2,\  \ldots, \delta_{n-1}-\delta_n,\ \delta_n-\epsilon_1,\ \epsilon_1-\epsilon_2,\ \ldots, \epsilon_{m-1}-\epsilon_m,\ \epsilon_m. 
\label{osp2m12n}
\end{equation}
Also in this case there exists a useful $\mathbb{Z}$-gradation of $\mathfrak{g}= \mathfrak{osp}(2m+1|2n)$: 
$\mathfrak{g}=\mathfrak{g}_{-2}\oplus\mathfrak{g}_{-1}\oplus\mathfrak{g}_0\oplus\mathfrak{g}_{+1}\oplus\mathfrak{g}_{+2}$, where each
$\mathfrak{g}_j$ is spanned by the root vectors corresponding to the following roots:
\[
\begin{array}{ccccc}
\phantom{mm}\mathfrak{g}_{-2}\phantom{mm} &\phantom{mm}\mathfrak{g}_{-1}\phantom{mm} &\phantom{mm}\mathfrak{g}_{0}\phantom{mm} &\phantom{mm}\mathfrak{g}_{+1}\phantom{mm} &\phantom{mm}\mathfrak{g}_{+2}\phantom{mm} \\
-\delta_i-\delta_j & -\delta_i & \delta_i-\delta_j & \delta_i & \delta_i+\delta_j\\
-\epsilon_i-\epsilon_j\ (i\ne j) & -\epsilon_i & \epsilon_i-\epsilon_j & \epsilon_i & \epsilon_i+\epsilon_j\ (i\ne j)\\
-\epsilon_i-\delta_j & & \pm(\epsilon_i-\delta_j) & & \epsilon_i+\delta_j
\end{array}
\]
So in this case $\mathfrak{g}_{0}=\mathfrak{gl}(m|n)$. Note that this $\mathbb{Z}$-gradation is different from the more
common $\mathbb{Z}$-gradation for which $\mathfrak{g}_{0}=\mathfrak{g}_{\bar 0}$~\cite{dict,Gould}.

Let us now consider the irreducible highest weight representation with highest weight given by
$(\frac{p}{2},\ldots,\frac{p}{2};-\frac{p}{2},\ldots, -\frac{p}{2})= \sum_{i=1}^m \frac{p}{2}\epsilon_i + \sum_{i=1}^n (-\frac{p}{2})\delta_i$, where $p$ is a non-negative integer. 
This representation will be denoted by $V=V_{\mathfrak{osp}(2m+1|2n)}^{(p/2)^m,(-p/2)^n}$.
According to the simple roots system~\eqref{osp2m12n}, its Dynkin labels are $[0,0,\ldots,0,p]$.
The action of $\mathfrak{g}_0=\mathfrak{gl}(m|n)$ on the highest weight vector $v$ is again trivial, so the $\mathbb{Z}$-gradation of
$\mathfrak{g}$ induces a $\mathbb{Z}$-gradation of $V$ with $\dim V_0=1$, and one can again study $\dim_t(V)$ and $\sdim_t(V)$.
In the current situation, the structure of the irreducible highest weight representation with highest weight 
$(\frac{p}{2},\ldots,\frac{p}{2};-\frac{p}{2},\ldots, -\frac{p}{2})$
is completely the same as the irreducible lowest weight representation with lowest weight 
$(-\frac{p}{2},\ldots,-\frac{p}{2};\frac{p}{2},\ldots, \frac{p}{2})$.
This representation has been studied recently and its structure is well known: it is the parastatistics representation of order $p$~\cite{parast}.
Using the notation $x_i=\hbox{e}^{-\epsilon_i}$, $y_i=\hbox{e}^{-\delta_i}$, the following character formula holds:
\begin{equation}
\ch V_{\mathfrak{osp}(2m+1|2n)}^{(p/2)^m,(-p/2)^n} = (y_1\cdots y_n/x_1\cdots x_m)^{p/2} \sum_{\lambda,\ \lambda_1\leq p} s_\lambda(x|y).
\label{char-Bmn}
\end{equation}
So here the sum is over all partitions $\lambda$ inside the $(m,n)$-hook (otherwise $s_\lambda(x|y)$ is zero anyway) with 
$\lambda_1\leq p$, or equivalently $\ell(\lambda')\leq p$.
There is also an alternative character formula~\cite[Theorem~7]{parast}, but we do not need it here.

In order to determine $\sdim_t V_{\mathfrak{osp}(2m+1|2n)}^{(p/2)^m,(-p/2)^n}$, one should (apart from the factor in front of the above
sum) specify $x_i=t$ and $y_j=-t$ in the above character, and so one finds
\begin{align}
\sdim_t V_{\mathfrak{osp}(2m+1|2n)}^{(p/2)^m,(-p/2)^n} &= \sum_{\lambda,\ \lambda_1\leq p} s_\lambda(t,\ldots,t|-t,\ldots,-t) \nonumber\\
&= \sum_{\lambda,\ \lambda_1\leq p} s_\lambda(1,\ldots,1|-1,\ldots,-1)\, t^{|\lambda|}  \nonumber\\
&= \sum_{\lambda,\ \lambda_1\leq p} \sdim V_{\mathfrak{gl}(m|n)}^{\lambda} \, t^{|\lambda|}. 
\label{sdim-Bmn}
\end{align}
We can now make use of the expressions and properties of $\mathfrak{gl}(m|n)$ superdimensions, given at the end of section~\ref{sec2},
and thus specify the following three cases.

\vskip 2mm
\noindent {\bf Case 1: $m=n$, $\mathfrak{osp}(2n+1|2n)$.}\\
This is the simplest case, since all superdimensions of covariant representations of $\mathfrak{gl}(n|n)$ are zero, except when $\lambda=(0)$. Hence:
\begin{equation}
\sdim_t V_{\mathfrak{osp}(2n+1|2n)}^{(p/2)^n,(-p/2)^n} = 1.
\end{equation}

\vskip 2mm
\noindent {\bf Case 2: $m=n+k$, $\mathfrak{osp}(2n+2k+1|2n)$.}\\
Now it follows directly from~\eqref{sdim-Bmn} and~\eqref{dim-sdim} that
\begin{equation}
\sdim_t V_{\mathfrak{osp}(2m+1|2n)}^{(p/2)^m,(-p/2)^n} = \sum_{\lambda,\ \lambda_1\leq p} \dim V_{\mathfrak{gl}(k)}^{\lambda} \, t^{|\lambda|}
= \sum_{\lambda,\ \lambda_1\leq p,\ \ell(\lambda)\leq k} \dim V_{\mathfrak{gl}(k)}^{\lambda} \, t^{|\lambda|}.
\label{sdim-Bk1}
\end{equation}
So the original infinite sum reduces to a finite sum, running over all partitions $\lambda$ with 
$\lambda_1\leq p$ and $\ell(\lambda)\leq k$. In other words, the sum is over all partitions $\lambda$ whose Young diagrams fit
inside the $(k\times p)$ rectangle. 
However, such an expression is known, see~\cite{parafermion} for a detailed investigation (although the origin goes back to work 
of Bracken and Green~\cite{BG1}). 
Indeed, in the branching $\mathfrak{so}(2k+1) \supset \mathfrak{gl}(k)$, the $\mathfrak{so}(2k+1)$ highest weight representation
$V_{\mathfrak{so}(2k+1)}^{(p/2)^k}$ with highest weight $(\frac{p}{2},\ldots,\frac{p}{2})$ decomposes into the direct sum of 
covariant $\mathfrak{gl}(k)$ representation $V_{\mathfrak{gl}(k)}^{\lambda}$ with $\ell(\lambda')\leq p$.
Thus we find:
\begin{equation}
\sdim_t V_{\mathfrak{osp}(2m+1|2n)}^{(p/2)^m,(-p/2)^n} = \dim_t V_{\mathfrak{so}(2k+1)}^{(p/2)^k},
\label{sdim-Bk2}
\end{equation}
or, in terms of Dynkin labels:
\begin{equation}
\sdim_t [0,0,\ldots,0,p]_{\mathfrak{osp}(2n+2k+1|2n)} = \dim_t [0,\ldots,0,p]_{\mathfrak{so}(2k+1)}.
\label{sdim-Bk3}
\end{equation}

Let us give a few examples. For $k=3$ and $p=1$, one finds
\[
\sdim_t [0,0,\ldots,0,1]_{\mathfrak{osp}(2n+7|2n)} = \dim_t [0,0,1]_{\mathfrak{so}(7)}
=1+3t+3t^2+t^3.
\]
and thus $\sdim [0,0,\ldots,0,1]_{\mathfrak{osp}(2n+7|2n)} = \dim [0,0,1]_{\mathfrak{so}(7)} =8$.
For $k=3$ and $p=2$, one gets
\[
\sdim_t [0,0,\ldots,0,2]_{\mathfrak{osp}(2n+7|2n)} = \dim_t [0,0,2]_{\mathfrak{so}(7)}
=1+3t+9t^2+9t^3+9t^4+3t^5+t^6.
\]
and $\sdim [0,0,\ldots,0,2]_{\mathfrak{osp}(2n+7|2n)} = \dim [0,0,2]_{\mathfrak{so}(7)} =35$.

\vskip 2mm
\noindent {\bf Case 3: $n=m+k$, $\mathfrak{osp}(2m+1|2m+2k)$.}\\
Again following~\eqref{sdim-Bmn} and~\eqref{dim-sdim}, one finds
\begin{equation}
\sdim_t V_{\mathfrak{osp}(2m+1|2n)}^{(p/2)^m,(-p/2)^n} = \sum_{\lambda,\ \lambda_1\leq p,\ \lambda_1\leq k}
(-1)^{|\lambda|} \dim V_{\mathfrak{gl}(k)}^{\lambda'} \, t^{|\lambda|}
= \sum_{\mu,\ \ell(\mu)\leq\min(p,k)} \dim V_{\mathfrak{gl}(k)}^{\mu} \, (-t)^{|\mu|}.
\label{sdim-sBk1}
\end{equation}
The right hand side is the same expression as~\eqref{tdim-osp1} (with $t\rightarrow -t$).
Following~\eqref{tdim-osp2} one can write
\begin{equation}
\sdim_t V_{\mathfrak{osp}(2m+1|2n)}^{(p/2)^m,(-p/2)^n} 
=\dim_{-t} V_{\mathfrak{osp}(1|2k)}^{(-p/2)^k} = \frac{E_{k,p}(-t,\ldots,-t)}{(1+t)^k (1-t^2)^{k(k-1)/2}}.
\label{sdim-sBk2}
\end{equation}
Herein, a finite expression for $E_{k,p}(-t,\ldots,-t)$ can be found in~\eqref{Enpt}.
Note that in terms of Dynkin labels one could write:
\begin{equation}
\sdim_t [0,0,\ldots,0,p]_{\mathfrak{osp}(2m+1|2m+2k)} = \dim_{-t} [0,\ldots,0,-p]_{\mathfrak{osp}(1|2k)}.
\label{sdim-sBk3}
\end{equation}

Let us again consider some examples.
For $k=3$ and $p=1$, one finds
\[
\sdim_t [0,0,\ldots,0,1]_{\mathfrak{osp}(2m+1|2m+6)} = \dim_{-t} [0,0,-1]_{\mathfrak{osp}(1|6)}
=\frac{1}{(1+t)^3}=1-3t+6t^2-10t^3+15t^4-\cdots.
\]
One could say that the actual superdimension (for $t=1$) is $1/8$.
More general, for arbitrary $k$,
\[
\sdim_t [0,0,\ldots,0,1]_{\mathfrak{osp}(2m+1|2m+2k)} = \dim_{-t} [0,\ldots,0,-1]_{\mathfrak{osp}(1|2k)}
=\frac{1}{(1+t)^k}.
\]
For $k=3$ and $p=2$, one gets
\begin{align*}
& \sdim_t [0,0,\ldots,0,2]_{\mathfrak{osp}(2m+1|2m+6)} 
= \dim_{-t} [0,0,-2]_{\mathfrak{osp}(1|6)} \\
& \qquad = \frac{1+t^3}{(1+t)^3(1-t^2)^3} = 1-3t+9t^2-18t^3+36t^4-\cdots .
\end{align*}

\section{The Lie superalgebra $\mathfrak{osp}(2m|2n)$}
\label{sec5}

The second main case treated here is the Lie superalgebra $D(m,n)=\mathfrak{osp}(2m|2n)$.
The weight space of $\mathfrak{osp}(2m|2n)$ has the same basis $\epsilon_1,\ldots,\epsilon_m,\delta_1,\ldots,\delta_n$, and the distinguished set
of simple roots is now
\begin{equation}
\delta_1-\delta_2,\  \ldots, \delta_{n-1}-\delta_n,\ \delta_n-\epsilon_1,\ \epsilon_1-\epsilon_2,\ \ldots, \epsilon_{m-2}-\epsilon_{m-1},\ \epsilon_{m-1}-\epsilon_m,\ \epsilon_{m-1}+\epsilon_m. 
\label{osp2m2n}
\end{equation}
It will be helpful to see $D(m,n)$ as a subalgebra of $B(m,n)$. 
In fact, using the $\mathbb{Z}$-gradation 
$\mathfrak{g}_{-2}\oplus\mathfrak{g}_{-1}\oplus\mathfrak{g}_0\oplus\mathfrak{g}_{+1}\oplus\mathfrak{g}_{+2}$ 
of $\mathfrak{g}=\mathfrak{osp}(2m+1|2n)$ introduced in the previous section, 
it is easy to see that $\mathfrak{osp}(2m|2n) = \mathfrak{g}_{-2}\oplus\mathfrak{g}_0\oplus\mathfrak{g}_{+2}$, with root structure
\[
\begin{array}{ccc}
\phantom{mm}\mathfrak{g}_{-2}\phantom{mm}  &\phantom{mm}\mathfrak{g}_{0}\phantom{mm} &\phantom{mm}\mathfrak{g}_{+2}\phantom{mm} \\
-\delta_i-\delta_j & \delta_i-\delta_j & \delta_i+\delta_j\\
-\epsilon_i-\epsilon_j\ (i\ne j) & \epsilon_i-\epsilon_j & \epsilon_i+\epsilon_j\ (i\ne j)\\
-\epsilon_i-\delta_j & \pm(\epsilon_i-\delta_j) & \epsilon_i+\delta_j
\end{array}
\]
So also in this case $\mathfrak{g}_{0}=\mathfrak{gl}(m|n)$, and it will be useful to consider the 
representation of the previous section according to $\mathfrak{osp}(2m+1|2n)\supset\mathfrak{osp}(2m|2n)\supset\mathfrak{gl}(m|n)$.

The purpose is to study the irreducible highest weight representation of $\mathfrak{osp}(2m|2n)$ with highest weight given by
$(\frac{p}{2},\ldots,\frac{p}{2};-\frac{p}{2},\ldots, -\frac{p}{2})= \sum_{i=1}^m \frac{p}{2}\epsilon_i + \sum_{i=1}^n (-\frac{p}{2})\delta_i$, where $p$ is a non-negative integer, and denoted by $V=V_{\mathfrak{osp}(2m|2n)}^{(p/2)^m,(-p/2)^n}$.
According to the simple roots system~\eqref{osp2m2n}, its Dynkin labels are $[0,0,\ldots,0,p]$.

Contrary to $V_{\mathfrak{osp}(2m+1|2n)}^{(p/2)^m,(-p/2)^n}$, $V_{\mathfrak{osp}(2m|2n)}^{(p/2)^m,(-p/2)^n}$ has not (yet) been studied in the context of parastatistics.
However, the techniques of~\cite{parast} can also be used in the current case.
In particular, let $v$ be the vector of highest weight $(\frac{p}{2},\ldots,\frac{p}{2};-\frac{p}{2},\ldots, -\frac{p}{2})$, 
then $\mathbb{C}v$ is a trivial one-dimensional module of $\mathfrak{g}_0\oplus\mathfrak{g}_{+2}$, and we consider the 
induced module
\[
{\overline V}_{\mathfrak{osp}(2m|2n)}^{(p/2)^m,(-p/2)^n}= 
\hbox{Ind}_{\mathfrak{g}_0\oplus\mathfrak{g}_{+2}}^{\mathfrak{osp}(2m|2n)} \mathbb{C}v.
\]
Using the notation $x_i=\hbox{e}^{-\epsilon_i}$, $y_i=\hbox{e}^{-\delta_i}$, and the above root structure, the character is given by:
\begin{equation}
\ch {\overline V}_{\mathfrak{osp}(2m|2n)}^{(p/2)^m,(-p/2)^n} = (y_1\cdots y_n/x_1\cdots x_m)^{p/2} 
\frac{\prod_{i,j} (1+x_iy_j)}{\prod_{i<j}(1-x_ix_j)\prod_{i\leq j} (1-y_iy_j)}.
\label{char-Verma}
\end{equation}
Cummins and King~\cite{Cummins,CumminsKing} obtained an expansion of the above product in terms of supersymmetric Schur functions:
\begin{align}
& \frac{\prod_{i,j} (1+x_iy_j)}{\prod_{i<j}(1-x_ix_j)\prod_{i\leq j} (1-y_iy_j)} =
\sum_{\beta} s_\beta(x|y) \nonumber\\
& \qquad = 1+s_{1,1}(x|y)+s_{2,2}(x|y)+s_{1,1,1,1}(x|y)+s_{3,3}(x|y)+s_{2,2,1,1}(x|y)+s_{1,1,1,1,1,1}(x|y)+\cdots ,
\end{align}
i.e.\ the sum is over all partitions $\beta$ such that each part of $\beta$ appears twice.
This can also be written as:
\begin{equation}
\frac{\prod_{i,j} (1+x_iy_j)}{\prod_{i<j}(1-x_ix_j)\prod_{i\leq j} (1-y_iy_j)} =
\sum_{\delta} s_{\delta'}(x|y),
\end{equation}
where the sum is now over all partitions $\delta$ with even parts only.
Let us denote by ${\cal B}$ the set of partitions for which each part appears twice (including the zero partition), and by 
${\cal D}$ the set of partitions with even parts only.
So
\begin{equation}
\ch {\overline V}_{\mathfrak{osp}(2m|2n)}^{(p/2)^m,(-p/2)^n} = (y_1\cdots y_n/x_1\cdots x_m)^{p/2} 
\sum_{\lambda\in{\cal B}} s_\lambda(x|y).
\label{char-Verma1}
\end{equation}

Now we return to the irreducible representation $V_{\mathfrak{osp}(2m|2n)}^{(p/2)^m,(-p/2)^n}$.
On the one hand, $V_{\mathfrak{osp}(2m|2n)}^{(p/2)^m,(-p/2)^n}$ is a quotient module of 
${\overline V}_{\mathfrak{osp}(2m|2n)}^{(p/2)^m,(-p/2)^n}$ with character~\eqref{char-Verma1}.
On the other hand, $V_{\mathfrak{osp}(2m|2n)}^{(p/2)^m,(-p/2)^n}$ is a submodule
of $V_{\mathfrak{osp}(2m+1|2n)}^{(p/2)^m,(-p/2)^n}$ in the decomposition $\mathfrak{osp}(2m+1|2n) \supset \mathfrak{osp}(2m|2n)$,
with character~\eqref{char-Bmn}.
From this observation, one can deduce:
\begin{equation}
\ch V_{\mathfrak{osp}(2m|2n)}^{(p/2)^m,(-p/2)^n} = (y_1\cdots y_n/x_1\cdots x_m)^{p/2} 
\sum_{\lambda\in{\cal B},\ \lambda_1\leq p} s_\lambda(x|y).
\label{char-Dmn}
\end{equation}

To determine $\sdim_t V_{\mathfrak{osp}(2m|2n)}^{(p/2)^m,(-p/2)^n}$, one should (apart from the factor in front of the above
sum) specify again $x_i=t$ and $y_j=-t$ in the above character, and so one finds
\begin{align}
\sdim_t V_{\mathfrak{osp}(2m|2n)}^{(p/2)^m,(-p/2)^n} &= \sum_{\lambda\in{\cal B},\ \lambda_1\leq p} s_\lambda(t,\ldots,t|-t,\ldots,-t) \nonumber\\
&= \sum_{\lambda\in{\cal B},\ \lambda_1\leq p} s_\lambda(1,\ldots,1|-1,\ldots,-1)\, t^{|\lambda|}  \nonumber\\
&= \sum_{\lambda\in{\cal B},\ \lambda_1\leq p} \sdim V_{\mathfrak{gl}(m|n)}^{\lambda} \, t^{|\lambda|}. 
\label{sdim-Dmn}
\end{align}
As in the previous section, we now make use of properties of $\mathfrak{gl}(m|n)$ superdimensions.

\vskip 2mm
\noindent {\bf Case 1: $m=n$, $\mathfrak{osp}(2n|2n)$.}
\begin{equation}
\sdim_t V_{\mathfrak{osp}(2n|2n)}^{(p/2)^n,(-p/2)^n} = 1.
\end{equation}

\vskip 2mm
\noindent {\bf Case 2: $m=n+k$, $\mathfrak{osp}(2n+2k|2n)$.}\\
It follows directly from~\eqref{sdim-Dmn} and~\eqref{dim-sdim} that
\begin{equation}
\sdim_t V_{\mathfrak{osp}(2m|2n)}^{(p/2)^m,(-p/2)^n} = \sum_{\lambda\in{\cal B},\ \lambda_1\leq p} \dim V_{\mathfrak{gl}(k)}^{\lambda} \, t^{|\lambda|}
= \sum_{\lambda\in{\cal B},\ \lambda_1\leq p,\ \ell(\lambda)\leq k} \dim V_{\mathfrak{gl}(k)}^{\lambda} \, t^{|\lambda|}.
\label{sdim-Dk1}
\end{equation}
This is a finite sum. 
The summation is over all partitions $\lambda$ with each part appearing twice and whose Young diagrams fit
inside the $(k\times p)$ rectangle. 
Using a character formula for $\mathfrak{so}(2k)$, it is easy to see that this corresponds to the character of
an irreducible $\mathfrak{so}(2k)$ representation (see Appendix).
There is a distinction between the cases $k$ even and $k$ odd.
We find:
\begin{equation}
\sdim_t V_{\mathfrak{osp}(2m|2n)}^{(p/2)^m,(-p/2)^n} = \left\{
\begin{array}{ll}
\displaystyle \dim_t V_{\mathfrak{so}(2k)}^{(p/2)^{k}} & \hbox{ for }k\hbox{ even}, \\
\displaystyle \dim_t V_{\mathfrak{so}(2k)}^{(p/2)^{k-1},-p/2} & \hbox{ for }k\hbox{ odd}; 
\end{array} \right.
\label{sdim-Dk2}
\end{equation}
or, in terms of Dynkin labels:
\begin{equation}
\sdim_t [0,0,\ldots,0,p]_{\mathfrak{osp}(2n+2k|2n)} = \left\{
\begin{array}{ll}
\displaystyle \dim_t [0,\ldots,0,0,p]_{\mathfrak{so}(2k)} & \hbox{ for }k\hbox{ even}. \\
\displaystyle \dim_t [0,\ldots,0,p,0]_{\mathfrak{so}(2k)} & \hbox{ for }k\hbox{ odd}. 
\end{array} \right.
\label{sdim-Dk3}
\end{equation}

As an example, let $k=5$ and $p=1$:
\[
\sdim_t [0,0,\ldots,0,1]_{\mathfrak{osp}(2n+10|2n)} = \dim_t [0,0,0,1,0]_{\mathfrak{so}(10)}
=1+10t^2+5t^4,
\]
and thus $\sdim [0,0,\ldots,0,1]_{\mathfrak{osp}(2n+10|2n)} = \dim [0,0,0,1,0]_{\mathfrak{so}(10)} =16$.
For $k=5$ and $p=2$, one gets
\[
\sdim_t [0,0,\ldots,0,2]_{\mathfrak{osp}(2n+10|2n)} = \dim_t [0,0,0,2,0]_{\mathfrak{so}(10)}
=1+10t^2+55t^4+45t^6+15t^8,
\]
and $\sdim [0,0,\ldots,0,2]_{\mathfrak{osp}(2n+10|2n)} = \dim [0,0,0,2,0]_{\mathfrak{so}(10)} =126$.

\vskip 2mm
\noindent {\bf Case 3: $n=m+k$, $\mathfrak{osp}(2m|2m+2k)$.}\\
From~\eqref{sdim-Dmn} and~\eqref{dim-sdim} one obtains
\begin{align}
\sdim_t V_{\mathfrak{osp}(2m|2n)}^{(p/2)^m,(-p/2)^n} 
& = \sum_{\lambda\in{\cal B},\ \lambda_1\leq p} \dim V_{\mathfrak{gl}(k)}^{\lambda'} \, (-t)^{|\lambda|}
= \sum_{\lambda\in{\cal B},\ \lambda_1\leq \min(p,k)} \dim V_{\mathfrak{gl}(k)}^{\lambda'} \, (-t)^{|\lambda|} \nonumber\\
&= \sum_{\lambda\in{\cal D},\ \ell(\lambda)\leq \min(p,k)} \dim V_{\mathfrak{gl}(k)}^{\lambda} \, (-t)^{|\lambda|}.
\label{sdim-Dk31}
\end{align}
This is an infinite sum, over all partitions $\lambda$ with even parts only and with $\ell(\lambda)\leq \min(p,k)$.
Interestingly, this expression is known and follows from the character of the infinite-dimensional irreducible
$\mathfrak{sp}(2k)$ module with highest weight $(-\frac{p}{2},\ldots,-\frac{p}{2})$~\cite{King2013}.
Thus we find:
\begin{equation}
\sdim_t V_{\mathfrak{osp}(2m|2n)}^{(p/2)^m,(-p/2)^n} = \dim_{-t} V_{\mathfrak{sp}(2k)}^{(-p/2)^k},
\label{sdim-Dk32}
\end{equation}
or, in terms of Dynkin labels:
\begin{equation}
\sdim_t [0,0,\ldots,0,p]_{\mathfrak{osp}(2m|2m+2k)} = \dim_{-t} [0,\ldots,0,-p/2]_{\mathfrak{sp}(2k)}.
\label{sdim-Dk33}
\end{equation}

As an example, consider first $p=1$. The series of partitions appearing in~\eqref{sdim-Dk31} is
\[
(0)+(2)+(4)+(6)+(8)+\cdots.
\]
So, for $k=3$, one finds
\begin{align*}
& \sdim_t [0,0,\ldots,0,1]_{\mathfrak{osp}(2m|2m+6)} = \dim_{-t} [0,0,-1/2]_{\mathfrak{sp}(6)} \\
&\qquad =1+6t^2+15t^4+28t^6+45t^8+\cdots  = \frac{1}{2}(\frac{1}{(1+t)^3}+\frac{1}{(1-t)^3}).
\end{align*}
Also in general, one has
\[
\sdim_t [0,0,\ldots,0,1]_{\mathfrak{osp}(2m|2m+2k)} = \dim_{-t} [0,\ldots,0,-1/2]_{\mathfrak{sp}(2k)} 
= \frac{1}{2}(\frac{1}{(1+t)^k}+\frac{1}{(1-t)^k}).
\]
For $p=2$, the series of partitions~\eqref{sdim-Dk31} is
\[
(0)+(2)+(4)+(2,2)+(6)+(4,2)+(8)+(6,2)+(4,4)+\cdots.
\]
For $k=3$, one finds
\[
\sdim_t [0,0,\ldots,0,2]_{\mathfrak{osp}(2m|2m+6)} = \dim_{-t} [0,0,-1]_{\mathfrak{sp}(6)}
=1+6t^2+21t^4+55t^6+ 120t^8+\cdots  .
\]

\section{The Lie superalgebra $D(2,1;\alpha)$}
\label{sec6}

Among the families of simple orthosymplectic Lie superalgebras, there are besides $B(m,n)=\mathfrak{osp}(2m+1|2n)$
and $D(m,n)=\mathfrak{osp}(2m|2n)$, still $C(n+1)=\mathfrak{osp}(2|2n)$ and $D(2,1;\alpha)$.
Since $C(n+1)$ is a type~1 Lie superalgebra (in the sense of Kac~\cite{Kac}), 
there seem to be no representations that one could describe as ``spinor'' or ``self dual tensors''. 
So we shall not include $C(n+1)$ in this paper.
By the way, all finite-dimensional irreducible representations of $C(n+1)$ are known~\cite{VdJ-C}.

For the exceptional family $D(2,1;\alpha)$ (where $\alpha$ is a real parameter that can be chosen positive), 
there are representations that one could consider as spinors or self dual tensors.
The weight space of $D(2,1;\alpha)$ is usually expressed in the basis $\epsilon_1,\epsilon_2,\epsilon_3$, with the distinguished set
of simple roots given by
\begin{equation}
2\epsilon_2,\ \epsilon_1-\epsilon_2-\epsilon_3,\ 2\epsilon_3 . 
\label{D21}
\end{equation}
The purpose of this section is to consider irreducible representations $V(\Lambda)$ with highest weight $\Lambda$ given by the Dynkin labels $[0,0,p]$ in the above simple root system. 
These are infinite-dimensional representations.
For $p=1$ this is the ``spinor'' irrep of $D(2,1;\alpha)$, for $p=2$ the ``self dual tensor''.
In terms of the $\epsilon$-basis, the highest weight is given by $\Lambda=(-\frac{\alpha p}{\alpha+1},0,p)$.
Such representations are ``doubly atypical''.
All irreducible highest weight representations of $D(2,1;\alpha)$ have been described in~\cite{VdJ-D}.
For the irreps under consideration one can describe the structure by means of the decomposition with respect to the even subalgebra 
$\mathfrak{su}(1,1) \oplus \mathfrak{su}(2) \oplus \mathfrak{su}(2)$ (isomorphic to $\mathfrak{sp}(2)\oplus \mathfrak{so}(4)$).
From~\cite{VdJ-D}, the following decompositions hold:

$p=1$:
\begin{equation}
V(\Lambda) \rightarrow (-\frac{\alpha p}{\alpha+1};0,1) \oplus (-\frac{\alpha p}{\alpha+1}-1;1,0)
\label{p1}
\end{equation}

$p\geq 2$:
\begin{equation}
V(\Lambda) \rightarrow (-\frac{\alpha p}{\alpha+1};0,p) \oplus (-\frac{\alpha p}{\alpha+1}-1;1,p-1) \oplus (-\frac{\alpha p}{\alpha+1}-2;0,p-2).
\label{p2}
\end{equation}

The weights of such an $\mathfrak{su}(1,1)$ irrep with highest weight $\mu$ ($\mu<0$) are given by $(\mu,\mu-2,\mu-4,\ldots)$: this is a negative discrete series representation. 
So for the case $p=1$, with $\mu=-\frac{\alpha p}{\alpha+1}$, one has:
\begin{itemize}
\item at level $\mu$: the $\mathfrak{su}(2) \oplus \mathfrak{su}(2)$ irrep $(0,1)$ of dimension~2
\item at level $\mu-1$: the $\mathfrak{su}(2) \oplus \mathfrak{su}(2)$ irrep $(1,0)$ of dimension~2
\item at level $\mu-2$: the $\mathfrak{su}(2) \oplus \mathfrak{su}(2)$ irrep $(0,1)$ of dimension~2
\item at level $\mu-3$: the $\mathfrak{su}(2) \oplus \mathfrak{su}(2)$ irrep $(1,0)$ of dimension~2
\item etc.
\end{itemize}
Thus one could say that the superdimension of $V(\Lambda)$ according to the above gradation (i.e.\ according to the 
$\mathfrak{su}(1,1)$ diagonal element) is given by
\[
2-2t+2t^2-2t^3+2t^4-\cdots = \frac{2}{1+t}.
\]
In the limit $t\rightarrow 1$, this gives superdimension~1.

In the previous section, we always considered a gradation with respect to the subalgebra $\mathfrak{gl}(m|n)$. For the irrep under consideration,
the decomposition with respect to $D(2,1;\alpha) \rightarrow \mathfrak{gl}(2|1)$ consists of
\begin{itemize}
\item a $\mathfrak{gl}(2|1)$ singlet with weight $(\mu,0,1)$
\item a $\mathfrak{gl}(2|1)$ irrep with weights $(\mu,0,-1), (\mu-1,-1,0), (\mu-1,1,0), (\mu-2,0,1)$, of dimension~4 and superdimension~0
\item a $\mathfrak{gl}(2|1)$ irrep with weights $(\mu-2,0,-1), (\mu-3,-1,0), (\mu-3,1,0), (\mu-4,0,1)$, of dimension~4 and superdimension~0
\item etc.
\end{itemize}
Hence also according to this gradation the superdimension is $1+0+0+\cdots =1$, whereas the dimension is $1+4+4+\cdots$.

For the case $p\geq 2$, the decomposition is given by~\eqref{p2}. The dimensions of the 
$\mathfrak{su}(2) \oplus \mathfrak{su}(2)$ irreps appearing here are, respectively, $p+1$, $2p$ and $p-1$.
According to the gradation by the $\mathfrak{su}(1,1)$ diagonal element, one finds as superdimension:
\begin{align*}
& (p+1)-t(2p)+t^2((p+1)+(p-1))-t^3(2p)+t^4((p+1)+(p-1))+\cdots \\
& = (p+1)-2p t+2p t^2-2p t^3+2p t^4+\cdots = 1-p+\frac{2p}{1+t}.
\end{align*}
In the limit $t\rightarrow 1$, this gives superdimension~1.
Also according to the decomposition with respect to $\mathfrak{gl}(2|1)$ one finds a singlet followed by an infinite series of irreps of dimension~4 and superdimension~0, thus the same result.

Conclusion: all $D(2,1;\alpha)$ irreps $[0,0,p]$ have superdimension~1, which is the same as the dimension of the $\mathfrak{so}(2)$ irrep $[p]$.

\section{Summary: Dynkin diagrams}
\label{sec7}

The main results of sections~\ref{sec4} and~\ref{sec5} can be best summarized and illustrated by means of Dynkin diagrams, 
carrying the Kac-Dynkin labels of the representations studied here.

For the case $B(m,n)=\mathfrak{osp}(2m+1|2n)$, one has, in an obvious notation, from~\eqref{sdim-Bk3} and~\eqref{sdim-sBk3}:
\begin{align*}
& \sdim \left(\raisebox{-5mm}{\includegraphics{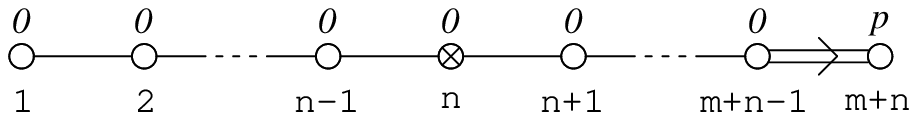}}\right) = \\
& \qquad \dim \left(\raisebox{-5mm}{\includegraphics{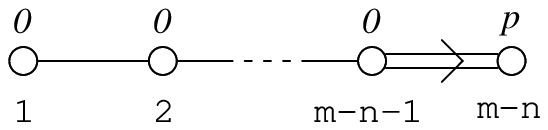}}\right) \qquad\hbox{ if } m>n\\
& \qquad \dim \left(\raisebox{-5mm}{\includegraphics{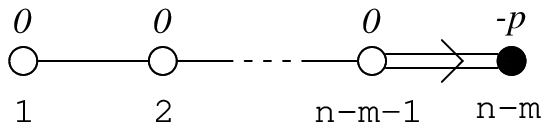}}\right) \qquad\hbox{ if } m<n\\
\end{align*}

For the case $D(m,n)=\mathfrak{osp}(2m|2n)$, the result is, following~\eqref{sdim-Dk3} and~\eqref{sdim-Dk33}:
\begin{align*}
& \sdim \left(\raisebox{-12mm}{\includegraphics{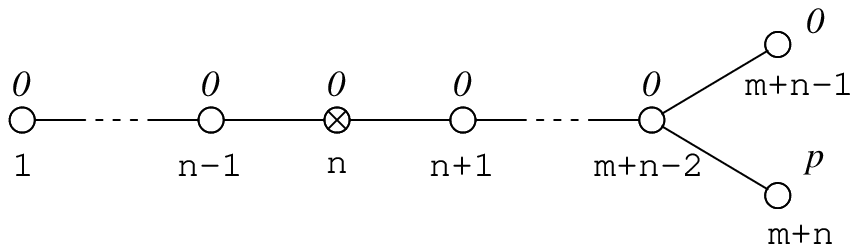}}\right) = \\
& \qquad \dim \left(\raisebox{-12mm}{\includegraphics{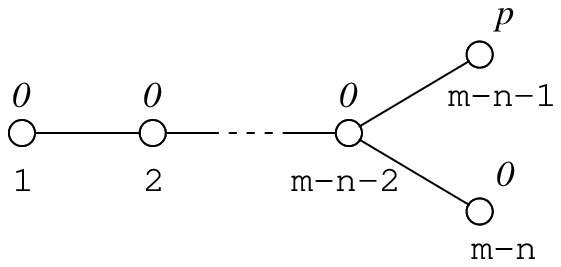}}\right) \qquad\hbox{ if } m>n \hbox{ and } m-n\hbox{ odd}\\
& \qquad \dim \left(\raisebox{-12mm}{\includegraphics{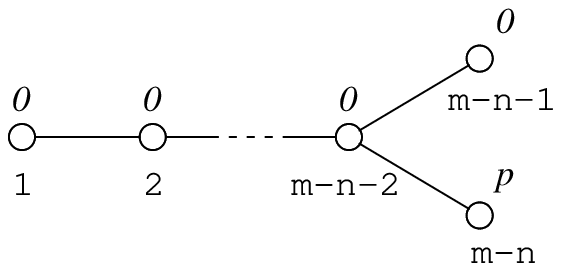}}\right) \qquad\hbox{ if } m>n\hbox{ and } m-n\hbox{ even}\\
& \qquad \dim \left(\raisebox{-5mm}{\includegraphics{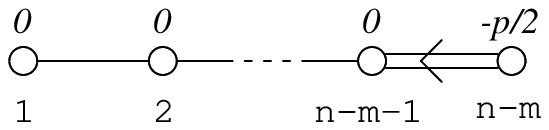}}\right) \qquad\hbox{ if } m<n\\
\end{align*}


\section*{Acknowledgments}
NIS  is thankful to  Professor  V.K. Dobrev for constructive discussions.	
NIS and JVdJ were supported by the Joint Research Project ``Lie superalgebras - applications in quantum theory'' in the framework of an international collaboration programme between the Research Foundation -- Flanders (FWO) and the Bulgarian Academy of Sciences. NIS was partially supported by Bulgarian National Science Fund, grant DFNI T02/6.
This research (JT-M) was supported in part by the Intramural Research Program of the NIH, U.S. National Library of Medicine.


\section*{Appendix}

In this appendix we give character formulas for the irreducible representations of $\mathfrak{so}(2k)$ which are relevant in the context of this paper.
Following the notation of this paper, these are the representations with Dynkin labels $[0,0,\ldots,0,0,p]$ and $[0,0,\ldots,0,p,0]$ (with $p$
a positive integer), or with highest weights $(\frac{p}{2},\frac{p}{2},\ldots,\frac{p}{2},\frac{p}{2})$ and $(\frac{p}{2},\frac{p}{2},\ldots,\frac{p}{2},-\frac{p}{2})$ respectively.
The modules will be denoted as $V_{\mathfrak{so}(2k)}^{(p/2)^k}$ and $V_{\mathfrak{so}(2k)}^{(p/2)^{k-1},-p/2}$.
Of course, the characters of these representations could be deduced from Weyl's formula.
Here, however, it is more useful to give these characters as an expansion in Schur functions.
This expression can actually be deduced from the $\mathfrak{so}(2k) \rightarrow \mathfrak{gl}(n)$ branching rule, given in~\cite[Section~IV]{King2000}.
Working out these branchings, there is a distinction between $k$ even and $k$ odd.
Explicitly, one finds:

\noindent {\bf Case $k$ even}:
\begin{equation}
\ch [0,\ldots,0,p]_{\mathfrak{so}(2k)} = \ch V_{\mathfrak{so}(2k)}^{(p/2)^{k}}
= (x_1\cdots x_k)^{-p/2} \sum_{\lambda \in {\cal B}:\ \lambda_1\leq p,\; \ell(\lambda)\leq k} s_\lambda (x).
\end{equation}
So the sum is over all partitions $\lambda$ for which each part appears twice, and such that the Young diagram of $\lambda$ fits inside the rectangle of width $p$ and height $k$.

\begin{equation}
\ch [0,\ldots,p,0]_{\mathfrak{so}(2k)} = \ch V_{\mathfrak{so}(2k)}^{(p/2)^{k-1},-p/2}
= (x_1\cdots x_k)^{-p/2} \sum_{\lambda \in {\cal B}:\ \lambda_1\leq p,\; \ell(\lambda)\leq k-2} s_{(p,\lambda)} (x).
\end{equation}
Here the sum is again over all partitions $\lambda$ for which each part appears twice, but such that the Young diagram of $\mu=(p,\lambda)$ fits inside the rectangle of width $p$ and height $k$.

\noindent {\bf Case $k$ odd}:
\begin{equation}
\ch [0,\ldots,0,p]_{\mathfrak{so}(2k)} = \ch V_{\mathfrak{so}(2k)}^{(p/2)^{k}}
= (x_1\cdots x_k)^{-p/2} \sum_{\lambda \in {\cal B}:\ \lambda_1\leq p,\; \ell(\lambda)\leq k-1} s_{(p,\lambda)} (x).
\end{equation}

\begin{equation}
\ch [0,\ldots,p,0]_{\mathfrak{so}(2k)} = \ch V_{\mathfrak{so}(2k)}^{(p/2)^{k-1},-p/2}
= (x_1\cdots x_k)^{-p/2} \sum_{\lambda \in {\cal B}:\ \lambda_1\leq p,\; \ell(\lambda)\leq k-1} s_{\lambda} (x).
\label{44}
\end{equation}

Note that in~\eqref{44}, due to the fact that $\lambda$ has an even number of parts (since $\lambda\in{\cal B}$), 
the condition $\ell(\lambda)\leq k-1$ could just as well be replaced by $\ell(\lambda)\leq k$ since $k$ is odd.

\end{document}